# An Integrated Air Quality Monitoring System for COVID-19 Virus Detection

Hanzhi Yang

University of Michigan

yanghz@umich.edu

**Abstract** – The pandemic of coronavirus disease 2019 (COVID-19) caused by syndrome-coronavirus-2 (SARS-CoV-2) has been found rapid and large-scale diagnosis to spread across the communities. The risk of infectious airborne aerosol transmission has been an area of increasing concern. In this regard, a selective optic detection of SARS-CoV-2 aerosol is highly desirable, which can actively collect the surrounding air samples and provide real-time feedbacks to users. In this work, we developed an integrated Air Quality Monitoring System (iAQMS) that enables real-time SARS-CoV-2 aerosol detection with high sensitivity. The detection device, inspired by localized surface plasmon resonance (LSPR), was designed to measure the intensity changes of light going through our biochips in which gold nanoparticles (AuNP) combined with antibodies were exposed to airflow containing virus aerosols. We found that the variation of the measured photocurrent ($\Delta I/I_0$) after 350 seconds of exposure to virus aerosols is a function of virus particle concentrations. By incorporating the air collection device and the optic detection device into a handheld sized device that can communicate with a smart-phone through Bluetooth serial, we demonstrate real-time and sensitive detection of SARS-CoV-2 virus aerosols with a concentration dynamic range of $10^{-5} \sim 10^{-1} \, pfu/\mu L$. Because of its sensitivity and reliability, our device can also be used for detection of other virus aerosols by simply switching the antibodies combined to the AuNP in the biochips.

## Introduction

The COVID-19 due to SARS-CoV-2 is currently causing a global pandemic. In USA, as of October 2021, there have been over 44 million confirmed cases and 0.72 million reported deaths [1]. All available evidence suggests that this virus spreads rapidly and of large-scale within and across the communities [2]. Regarding the transmission of SARS-CoV-2, the United States Centers for Disease Control and Prevention (CDC) has acknowledged that breathing in or touching eyes, noses, and mouths with small droplets and particles that contains the virus is one of the most common ways of getting infected by SARS-CoV-2 [3]. Aerosols, or droplet nuclei, are airborne particles that consist of components like fungi, pollen, bacteria, and viruses [4]. Early in 2003, aerosol transmission was reported by World Health Organization (WHO) to be responsible for a superspreading event of SARS in a housing block located in Hong Kong, China [5]. The WHO report identified that the virus aerosols were transported in the wastewater plumbing system in the building and then spread through the empty U-bends in bathrooms. Recent studies have shown that SARS-CoV-2 virus remained viable in aerosols for at least 3 hours with limited reduction of infectious titers [6]. To prevent the potential exposure to SARS-CoV-2 virus aerosols in certain enclosed spaces, a real-time airborne virus detector is needed.

In a previous work [7], a smartphone-based integrated microsystem for on-site collection and detection that enables real-time detection of indoor airborne microparticles with high sensitivity was developed. The device collected airborne microparticles by the Venturi effect. With consistent negative pressure generated by high-speed airflow, the collection device was shown to be able to collect microparticles regardless of their concentration in the air sample. The optimal design and operating conditions of the collection device were determined using finite element analysis (FEA) and demonstrated using a wide range of airborne microparticles of various densities in airflow. The collection device was integrated with a detection device consisting of a microfluidic particle trapping chamber and a complementary metal-oxide-semiconductor (CMOS) photodetector. The whole platform was operated under smartphone-based communication in an app created in Blynk IoT Platform (https://blynk.io/).

In this work, we upgraded the design of the previous device to have 3 channels of detection for different target particles (Fig. 1). Certain upgrades on mechanical design of the device enabled users to conveniently replace the biochips without needing to disassemble the whole device. We discussed how each of the design upgrades affected the detection performance for airborne microparticles. We then applied AuNP combining with antibodies to highly sensitive virus particle detection and showed the impact of agarose gel, AuNP, and light sources. We also created a standalone Android app for Internet of Things (IoT) to control the device and plot the real-time data from the sensors. We showed that the variation of measured photocurrent is a function of the concentration of potential aerosols containing viruses Finally, we demonstrated that the triple channel device is able to collect enough aerosols for detection with high sensitivity and fine consistency in a short time range and then send real-time notifications to users on the app.

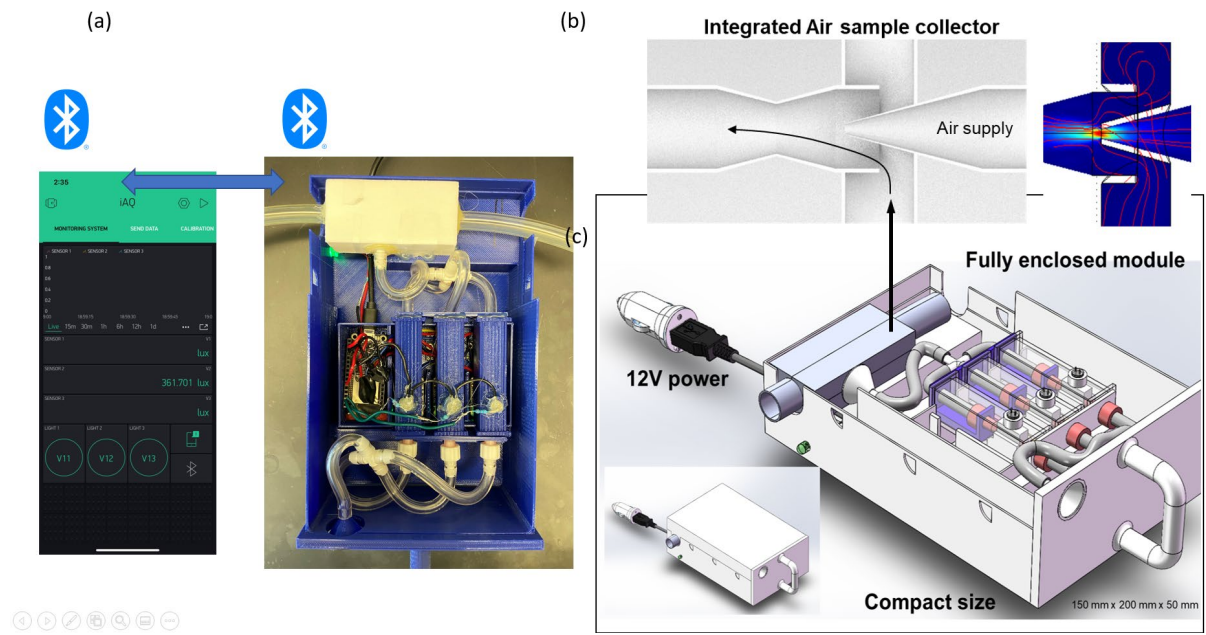

*Figure 1. Integrated Air Quality Monitoring System. (a) iAQMS is aimed to detect potential aerosols containing SARS-CoV-2 virus in an enclosed space such as a vehicle. The air quality is monitored using a smartphone-based Bluetooth communication. (b) The integrated air sample collector uses certain amount of air supply to generate a negative pressure that drags outside surrounding*

airflows into the detection device. The schematic diagram shows a simulation of airflow inside the collector [7]. (c) The entire platform integrates the air sample collector with a detection device that consists of 3 groups of light sources and photodetectors and a microcontroller with Bluetooth communication module embedded. The microcontroller receives commands from app on user's smartphone to start or stop the detection process and sends real-time detection results to smartphone on which they are shown as concurrently updating plots. The biochips in each channel contain antibodies for the target virus.

## Materials and methods

### i. Chemical materials

For conceptual design, we used different concentrations of methyl blue solutions to measure light intensities using the device. The methyl blue solutions were diluted into 10 mMol, 1 mMol, 0.1 mMol, 0.01 mMol, and 0.001 mMol. We used distilled water ($dH_2O$) and silicon dioxide ($SiO_2$) as control groups and tested SARS-CoV-2 inactivated virus in the final experiment. The initial concentration of the SARS-CoV-2 samples we have was 1 pfu/uL, and we diluted them into 0.1 pfu/uL, 0.01 pfu/uL, 0.001 pfu/uL, 0.0001 pfu/uL, and 0.00001 pfu/uL for concentration controls.

### ii. Fabrication of biochips

The biochip consists of 5 layers (Fig. 2(a)): (1) bottom layer which plays a role of adhesive layer and an optical window, (2) container layer which provides space for agarose gel and nanoprobe consisting of AuNP and antibodies, (3) middle layer functioning as a micro reaction chamber in which particles in airflow are collected and accumulated, (4) in/out layer which contains inlet and outlet for airflow needles, and (5) top layer functioning as an optical window and top enclosing layer. The biochip layers were cut out of methyl methacrylate (Acrylic) sheets by a laser cutter (Universal Laser Systems X2-600). The layers were then assembled using Gorilla Super Glue. To ensure the integrity of the assembled biochips, we pressed them using a paper clamp for 5 minutes. After the fabrication of all the layers, 100 uL of 5mg/mL agarose gel was injected into the reaction chamber of biochip, and then 10 uL of AuNP linked with antibodies was put onto the surface of agarose gel. Next, electrical tapes were wrapped around the biochip to achieve an enclosed reaction environment.

### iii. Fabrication of iAQMS

iAQMS is made up of 3 sections (Fig. 2(b)): (1) air sample collection section, (2) detection section, and (3) biochip loader section. All parts are designed in SolidWorks. Using a 3D printer (ALUNAR 3D Printer Prusa I3 Kit), we produced each part with polylactic acid (PLA). After printing out every part of iAQMS, we assembled the device and bonded some of the parts using hot glue. The tubes for airflow collection are TYRON R-3603 laboratory and vacuum tubing. All electronic components including microcontroller, multiplexer, and photodetector are purchased from Adafruit (https://www.adafruit.com/).

### iv. Integrated detector

A commercial CMOS photodetector (Adafruit TSL2591 light sensor) was connected to the esp32 (Adafruit esp32 feather) microcontroller via $I^2C$ communication protocol. The operation voltage of the photodetector and the logic voltage of esp32 microcontroller are both 3.3 V, so they are directly connected using jumper wires soldered on each pin. Since the smartphone app communicates with the device only through Bluetooth 2.0 serial, for power saving, in esp32 microcontroller settings, all wireless communication protocols other than Bluetooth classic such as Wi-Fi and Bluetooth Low Energy (BLE) are turned off. Then the light sources (Adafruit LED Sequins - Ruby Red, I = 50 mcd, $\lambda_p$ = 632 nm) were soldered onto microcontrollers and sticked onto the detection channels using hot glue. For the app controlling data communication and display, we used MIT App Inventor (https://appinventor.mit.edu/) to create a quick demo of the IoT app.

v.  Experiment setup

To proceed the detection of aerosols containing SARS-CoV-2 virus, we used a WH-2000 vaporizer to simulate an aerosol environment using water and virus samples. In the chamber of vaporizer, 3 mL of water and 60 uL of samples (water, SiO2, or SARS-CoV-2 virus) were put int and used to generate an enclosed environment filled up with aerosols. A tube with outer diameter (OD) of 3/8'' was connected to the chamber so that the aerosols can move through. The end of the tube was connected to the detection device by an OD 2mm needle inserted into the biochip, on the other side of which an OD 1.5mm needle were inserted and connected to the collection device. Once turned on, the vaporizer filled the chamber with "mist" of mixtures of water and sample particles, and the negative pressure generated by the collection device dragged the aerosols into the biochip, in which most of the particles accumulated and thus touched the AuNP with antibodies to form a reaction. Such reaction would result in a change in the absorbance of the medium including the AuNP and the agarose gel and could be detected by the CMOS photodetector beneath the biochip and shown on the app we developed (Fig. 2(c)).

vi.  Virus detection and LSPR

The detection of virus is inspired by LSPR which is an optical phenomenon caused by light interacting with conductive nanoparticles (NPs), e.g., AuNP, the sizes of which are smaller than the incident wavelength. For a surface plasmon resonance (SPR), the electric field of the light can be stored to collectively excite the electrons within a conduction band, resulting in coherent localized plasmon oscillations with a resonance frequency strongly depending on the composition, size, geometry, and separation distance of NPs [8]. When combined with virus on the S-protein, the plasmonic nanoprobe (AuNP with antibodies) would have its LSPR properties changed. Such change has been controlled to have a significant increase in absorbance under around 650 nm wavelength of light sources by the design of our AuNP. The definition of light absorbance is the logarithm of the ratio of incident to transmitted radiant power through a sample [9]:

$$A = \log_{10}(\frac{I_{in}}{I_{out}})$$ (eq. 1)

in which,

$A$ is the absorbance,

$I_{in}$ is the input light intensity, or in this work, the light intensity resulted from nanoprobe before detection,

$I_{out}$ is the output light intensity, i.e., the light intensity resulted from nanoprobe after exposed to virus.

However, due to the limitation of detection, such procedure was simplified to measuring the variation of the photocurrent, i.e., $\Delta I/I_0$, in which $\Delta I = I_0 - I$, $I$ is the real-time measured light intensity, and $I_0$ is the initial light intensity from nanoprobe before detection. Theoretically, the real-time photocurrent variation would gradually decrease at a decreasing rate until a steady state, and the decreasing rate is affected by the concentration of virus since it influences the rate of reaction with antibodies on AuNP. Therefore, after a certain time range of detection, the resulting photocurrent variation is related to the concentration of virus particles in surrounding environment, and that means, once a variation value of light intensity has been computed, the coherent virus particle concentration can be found.

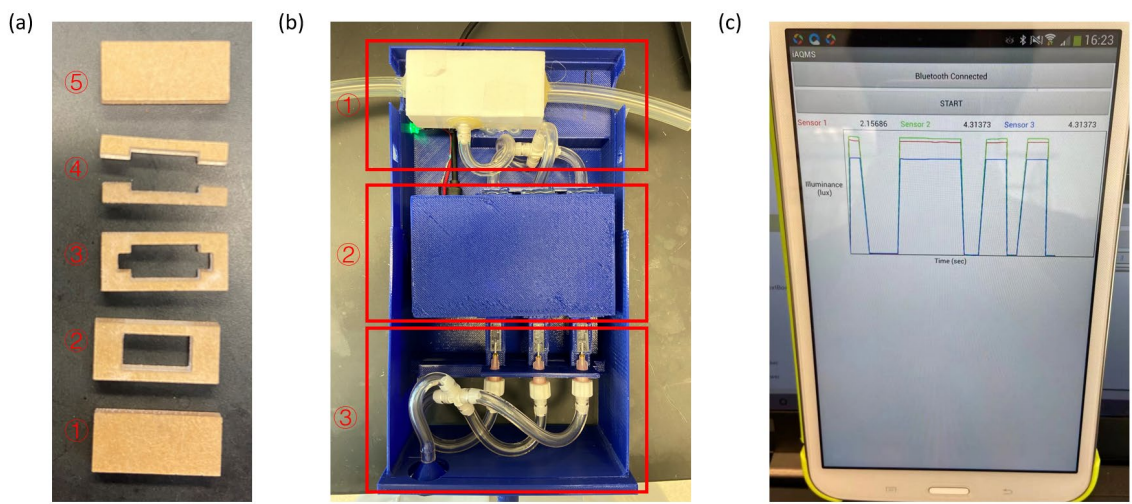

*Figure 2. Design of iAQMS device. (a) Biochip parts. (1) Bottom layer, (2) container layer, (3) middle layer, (4) in/out layer, and (5) top layer. (b) iAQMS device sections. (1) Collection section, (2) detection section, and (3) loading section. (c) Android app UI with sample sensor readings shown in the plot.*

## Results and discussion

i. Design of iAQMS detection section

The detection device consists of three light sources, one multiplexer, three photodetectors, three bridges, one sample holder, and one microcontroller unit. In this section, the multiplexer (Adafruit TCA9548A) is used to separate the addresses of the three photodetectors so that the microcontroller unit can communicate with them individually through I2C channel because the photodetectors we used on iAQMS have locked address and thus can not be altered. In its code, the microcontroller firstly recognizes the address of the multiplexer and from it gains a list of available sub-addresses that are connected to the photodetectors. Every time when the microcontroller reads the sensor values, it selects the sub-address of the target light sensor, proceeds the sensor reading and data processing programs, and then selects the next sub-address and repeats. Over the photodetectors, the bridge part holds the sample holder in which three biochips are placed. A LED light source is sticked by hot glue over the square hole on the bridge. In this case, the light beam from the LED goes through the square hole, and then the biochip, and eventually reaches the photodetector, which detects any changing behaviors of the light, indicating the chemical reactions occurring inside the biochip caused by virus particles linking with the nanoprobe. The needle used to transfer air along with aerosols is inserted into the bridge and poked inside of the biochip as shown in the figure. Such design follows the theories developed in the previous work [7] such that particles collected can accumulate inside the biochips and thus react with nanoprobes. In the next section, the impacts of light sources' wavelengths and the concentrations of the chemicals in biochip are examined.

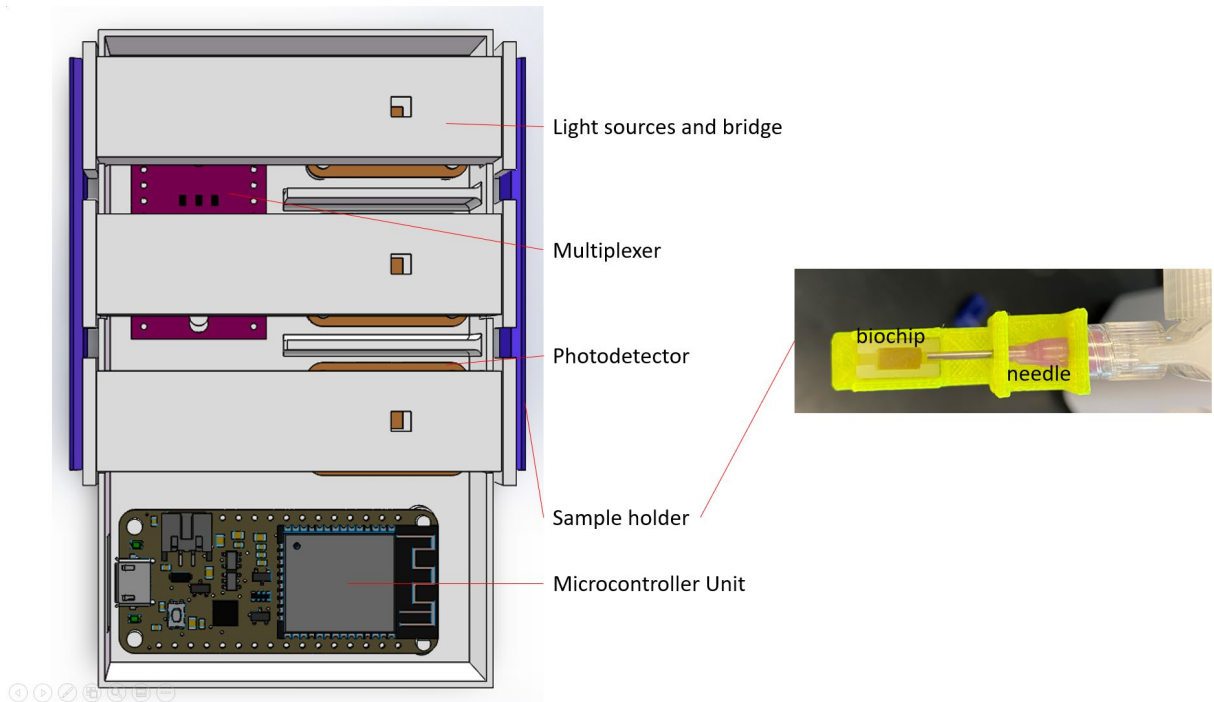

*Figure 3. Detection section inner view.*

ii.     Impacts of agarose gel, AuNP, and light source

Using SiO2 nanoparticles (SiO2 NP), we conducted experiments to examine the influence on sensitivity of the detection device by agarose gel, AuNP, and light sources. In figure 4(a), the absorbance due to SiO2 NP remains under 0.005 a.u regardless of the concentration of NP when no agarose gel nor AuNP are added in the biochip. Once certain amount of SiO2 NP is injected into the biochip, the detected SiO2 NP absorbance increases to 0.02 – 0.04 a.u for various concentrations of NP in figure 4(b). A significant increase in absorbance along the concentrations of SiO2 NP is discovered, shown in figure 4(c), when the light source is switched from white light with an average wavelength at approximately 470 nm to ruby red light with wavelength at 632 nm. Such increase in the sensitivity gets higher with AuNP added into the biochip. Note that in this experiment since no chemical reaction occurs in the biochip reaction chamber, the absorbance measured is resulted from SiO2 NP blocking the light from reaching the light sensor; hence, the large error shown in the plots is meaningless, and only the impact of agarose gel, AuNP, and light sources, which are essential elements in the iAQMS detection device, on the detection sensitivity is examined in this test. Following the results of this test, we concluded that for detecting particles collected in the biochip with high sensitivity, 100 uL of agarose gel, 10 uL of AuNP attached with particle detection probes (e.g., antibodies, proteins), and ruby red LED the wavelength of which is at constantly 632 nm are used in the iAQMS detection device.

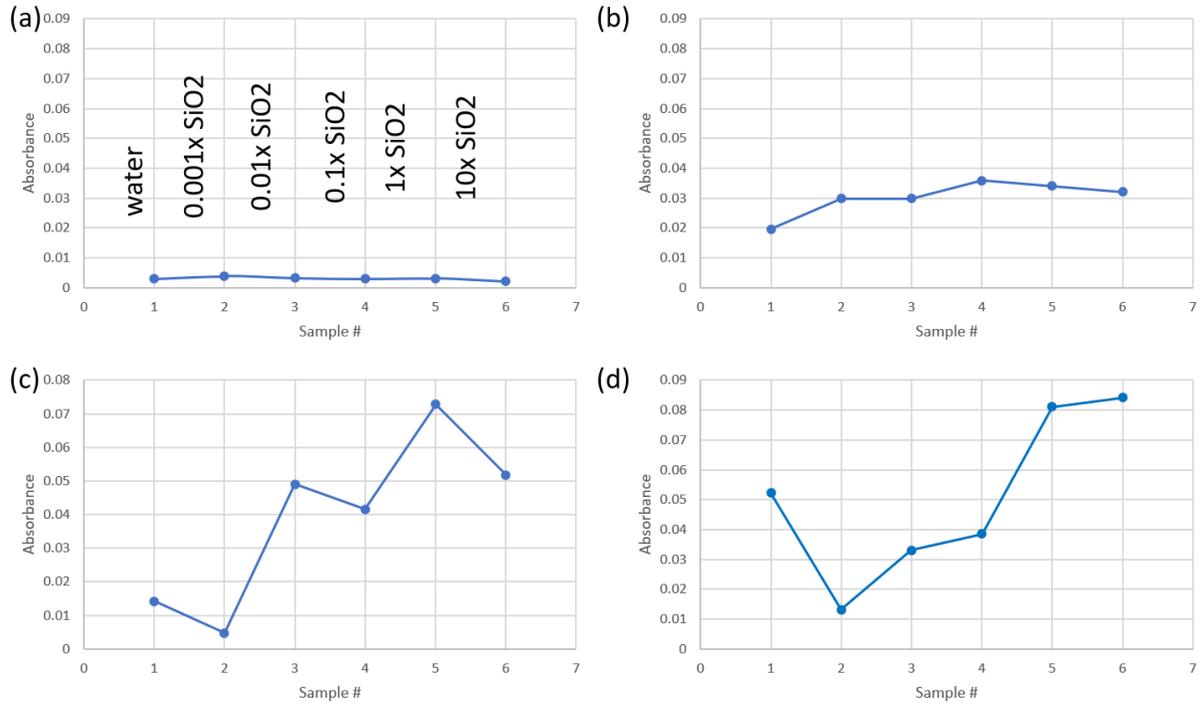

*Figure 4. Influence on light absorbance change by agarose gel, AuNP, and light source. (a) no agarose gel, no AuNP, white light source, (b) with agarose gel, no AuNP, white light source, (c) with agarose gel, no AuNP, red light source, (d) with agarose gel, with AuNP, red light source. Sample 1: distilled water, sample 2-6: 0.001x – 10x SiO2 NP.*

iii.     Detection of virus and bacteria

In this section, the ability of detecting virus and bacteria particles is examined on a single channel of detection. The detection process took 350 seconds to complete. The average sensor readings from the last 10 seconds were recorded and processed to the photocurrent variations. As shown in figure 5(a), the average photocurrent variations show a linear relationship to the concentration of virus particles, yielding

$$\frac{\Delta I}{I_0} = 0.38 + 0.046 \cdot \ln(C) \quad \text{(eq. 2)}$$

with an r-square value of 0.99. From the three times of testing, the samples of 1e-5 and 1e-3 pfu/uL concentrations have relatively large errors.

In figure 5(d), the average photocurrent variations versus the concentration of bacteria particles gives

$$\frac{\Delta I}{I_0} = 0.29 + 0.021 \cdot \ln(C) \quad \text{(eq. 3)}$$

with an r-square value of 0.97. For the bacteria tests, the samples of 1e-5 and 0.1 pfu/uL concentrations have quite large errors. The real-time records of variations of measured photocurrents by various concentrations of virus and bacteria samples are shown in figure 5(b) and figure 5(e). As shown, the noise level of virus particle detections is relatively low. However, the 0.0001 pfu/uL sample showed a negative photocurrent variation for 250 seconds during the detection, indicating that sometimes the detection may fail due to factors such as water droplets in the biochip, which might be caused by unstable air flow, and glue pollution on the surface of biochips, which could be an issue that need to be solved for biochip fabrications. For bacteria detection real-time record, the sensor readings show a much faster rate of getting to steady state as after around 100 seconds of detection the photocurrent reaches a balanced value with low deviation. This plot shows that the detection of bacteria particles still needs more calibrations to lower the noise level and increase the sensitivity for low concentration samples. To examine the consistency among all the 3 channels of detection, we applied the configurations of the single channel to all the three channels and proceeded tests using inactivated SARS-CoV-2 virus samples with a concentration of 0.1 pfu/uL. As shown in figure 5(c), the three detection channels reach a similar final value after 350 seconds of detecting, the photocurrent variation terms gradually increased from 0 to 0.20, which is slightly lower than the readings from single channel test. The reason of such decreased final reading is that the volume of virus particle sample initially placed in the vaporizer chamber was the same as that for single channel test, but in this case, the virus particles are distributed to three channels instead of one, making the amount of the potential particles "caught" by the nanoprobe inside biochip in three channels lower. As a result, for the same initial volume of virus particles, the three-channel detection gains a slightly lower final reading than a single channel detection.

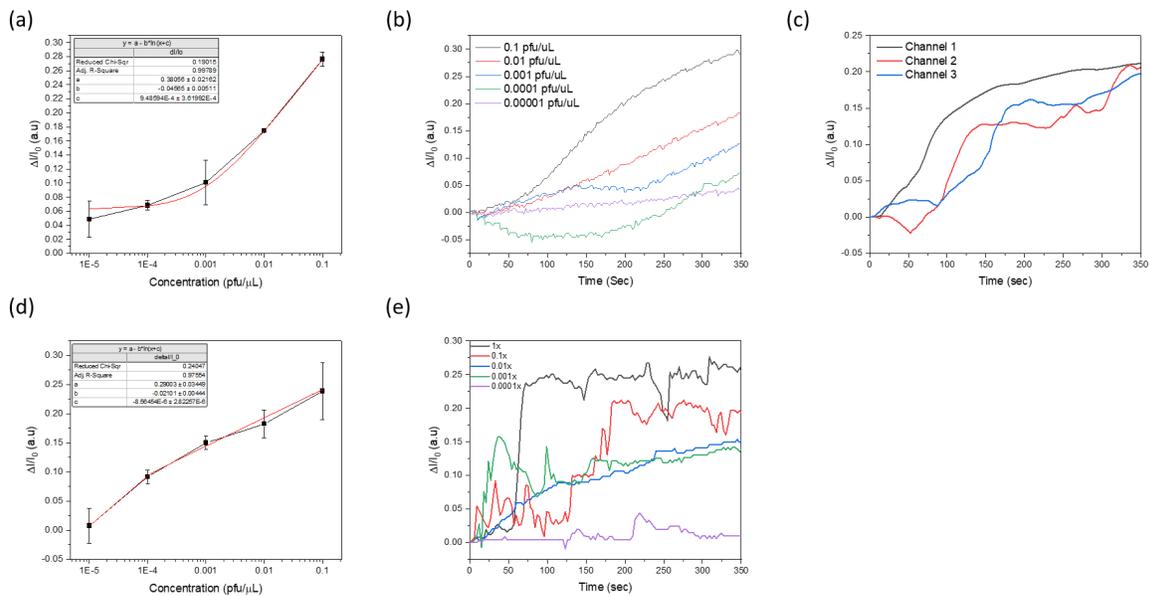

*Figure 5. Virus and bacteria detection test results. (a) Virus particles, (b) Virus particles real-time record, (c) Virus particles real-time record on three channels using 0.1 pfu/uL sample, (d) Bacteria particles, (e) Bacteria particles real-time record. (a) and (d) are conducted for three times and taken the averages and standard deviations.*

iv. Design of Android APP

To make the operation of iAQMS device more user friendly and mobile, we developed an Android app which communicates with the device through Bluetooth and therefore controls the light sources and receives and displays the real-time sensor readings on the smartphone screen. The reason of choosing Android as target operating system is that (1) Android system is open-sourced, so developers gain fully control on their apps in the system and (2) Android system is widely used not only on cellphones but also on many other devices like in-vehicle computer system, which is the target space to the iAQMS device. In this section, the evolution of the Android App from a prototype on Blynk to an independent Android app is presented.

Blynk is an open-sourced software required to prototype, deploy, and remotely manage connected electronic devices to form an IoT. It provides a lot of modules like buttons, plots, switches, etc. so that users can design and program the IoT app easily and quickly. Figure 6(a) shows the prototype app we created in Blynk. It consists of a plot in which real-time readings of all three light sensors are displayed. Below the plot, the numeric values of the sensor readings are shown. Users can also control the light sources on/off individually using the three buttons shown as V11, V12, and V13, beside which the control module of Bluetooth connection is placed. Blynk provides a convenient and easy platform to develop IoT applications, but a disadvantage of Blynk is that all apps using its platform, even though they are only using Bluetooth communication, need to have internet connection. This made Blynk a less desirable app development tool for our project because the requirement of internet connection makes it less convenient to operate the device when the internet connection is limited, e.g., in a car where the computer system is not connected online.

Therefore, MIT App Inventor was used to create a standalone Android app prototype. MIT App Inventor is a web application integrated development environment originally provided by Google, and now maintained by the Massachusetts Institute of Technology (MIT). It allows newcomers to computer programming to create apps of Android platform. MIT App Inventor, like Blynk, has many pre-programmed modules like buttons, textboxes, etc., for users to create UI fast. As shown in figure 6(b), the MIT App Inventor prototype recreates the UI in the Blynk app, including two buttons controlling the Bluetooth connection and the light sources, three textboxes showing the numeric values of three sensor readings, and a figure canvas plotting the real-time sensor readings. MIT App Inventor uses "drag-and-drop" block programming, making the coding for the app quite convenient. The logic of the app is that for every 100 ms of operating, it listens to the Bluetooth channel for updated data. Once an updated data is received in such string variable form as "sensor1 reading | sensor2 reading | sensor3 reading", the app splits the string into three parts: "sensor1 reading", "sensor2 reading", and "sensor3 reading" based on the location of the character "|". After turning the three strings above to float variables, the numbers can be shown in the textboxes and plotted in the figure. Although the MIT App Inventor app provides a clear logic of how the app runs, it has limited capability of rendering the UI and managing the Bluetooth connections.

As a result, for the beta version of iAQMS Android app, Android Studio was used. Android Studio is the official integrated development environment (IDE) for Google's Android operating system [10]. It provides a powerful tool for Android development including Android emulator environment, gradle-based build support, and a rich layout editor that allows users to drag-and-drop UI components and provides option to preview the layouts on multiple screen configurations. In Android studio, the plotting data function is achieved by the graphview module. This module simplifies the code of plotting data to one single command addDatapoint(x,y). As shown in figure 6(c), a prototype app developed in Android Studio uses a similar UI layout as that in the MIT App Inventor app. Note that for the current beta version development, the UI rendering has not been proceeded yet, so the current UI seems quite crude, but since Android Studio provides much flexibility of controlling and updating the app, for future work, a better app UI design would be accomplished.

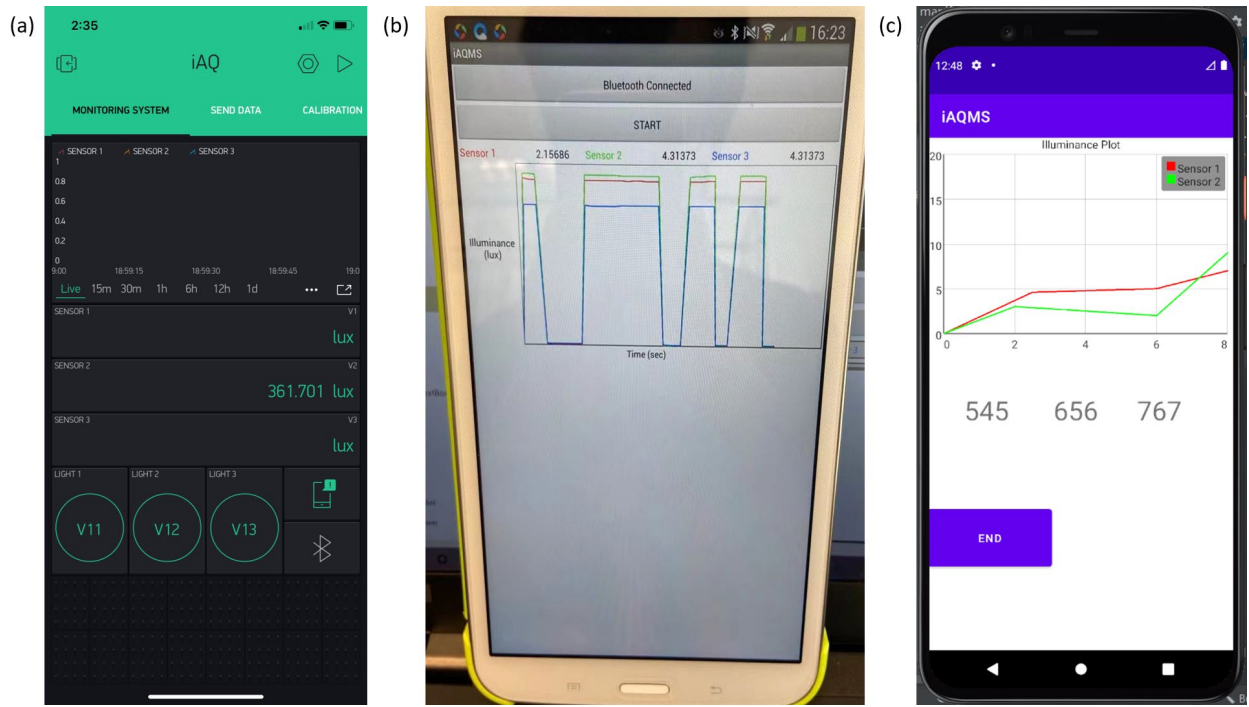

*Figure 6. iAQMS app. (a) Blynk IoT platform, (b) MIT App Inventor, (c) Android studio*

## Conclusions

In this work, we presented an integrated air quality monitoring system with rapid and sensitive aerosol collection and detection. The collection is inspired by Venturi effect, generating negative pressure that drags air flow containing aerosols into the biochips for detection. The detection involves the use of AuNP combined with antibody of SARS-CoV-2 virus as a nano-probe. By measuring the real-time changes in light intensity, the detection section expresses the potential concentration of virus particles in surrounding air as a function of measured variation of photocurrent. We developed an Android IoT app so that users can use to control the device through Bluetooth connection and view real-time detection feedbacks on a plot. We have open-sourced the app beta version and microcontroller unit program on GitHub: https://github.com/LesterYHZ/Integrated-Air-Quality-Monitoring-System-app.git. We demonstrated real-time and sensitive detection of SARS-CoV-2 virus aerosols with a concentration dynamic range of $10^{-5} \sim 10^{-1}\ pfu/\mu L$. Because of its sensitivity and reliability, our device can also be used for detection of other virus aerosols by simply switching the antibodies linked to the AuNP in the biochips.